\documentclass[pre,lengthcheck,showpacs,amssymb]{revtex4}
\usepackage{amsmath,bbm,epsfig}
\usepackage{amsfonts}
\usepackage{mathrsfs,psfrag}
\usepackage{graphicx,placeins}
\newcommand{\jpb}{J.\ Phys.\ B\ }
\newcommand{\epl}{Europhys.\ Lett.\ }

\newcommand{\jpa}{J.\ Phys.\ A\ }

\begin{document}

\title{Pseudo-classical theory for fidelity of nearly resonant quantum rotors}

\author{Martina Abb,$^1$ Italo Guarneri,$^{2}$ and Sandro Wimberger$^{1}$}

\affiliation{
$^1$Institut f\"ur Theoretische Physik, Universit\"at Heidelberg, Philosophenweg 19, 69120 Heidelberg, Germany \\
$^{2}$Center for Nonlinear and Complex Systems, Universit\`a dell'Insubria, Via Valleggio 11, 22100 Como; INFN, Sezione di Pavia, Via Bassi 6, 27100 Pavia, Italy
}

\date{\today}

\begin{abstract}

\noindent Using a semiclassical ansatz we analytically predict for the fidelity of $\delta$-kicked rotors the occurrence of revivals and the disappearance of intermediate revival peaks arising from the breaking of a symmetry in the initial conditions. A numerical verification of the predicted effects is given and experimental ramifications are discussed.

\end{abstract}


\maketitle


Besides entanglement in multipartite systems, it is the evolution of phases and the superposition principle which distinguishes a quantum from a classical system. Phase evolutions can be monitored in many ways, e.g., by correlation functions \cite{WM2008}. A quantity which has gained interest in the last decade is fidelity \cite{Gorin200633}, defined as the overlap of two wavefunctions subjected to slightly different temporal evolutions. The temporal evolution of this quantum fidelity crucially depends on evolving relative phases. For many-particle systems, fidelity can be viewed as a Hilbert space measure to study quantum phase transitions \cite{BV2007} and the regular-to-chaotic transition in complex quantum systems \cite{PLW2009}. For single-particle evolutions fidelity was measured in electromagnetic wave \cite{fid-micro} and matter wave \cite{andersen:104102} billiards, and with two different methods for periodically kicked cold atoms \cite{fid-ox,fid-har}.

The latter system is a realization of the quantum kicked rotor (QKR), the standard model for low-dimensional quantum chaos and the occurrence of dynamical localization \cite{Izrailev1990299}. Great interest in the QKR has reemerged in the study of its quantum resonant motion \cite{WGF03,WB2006,PhysRevLett.96.160403,PhysRevE.74.045201,PhysRevLett.94.174103,OC.179.137,PhysRevE.70.056206} and related accelerator modes \cite{PhysRevLett.83.4447,PhysRevLett.89.084101,PhysRevLett.96.164101,PhysRevLett.90.124102}. These two regimes are far from the classical limit of the QKR and, therefore, governed by distinct quantum effects. Nevertheless, close to quantum resonance the system can be described (pseudo-)classically with a new Planck's constant, which is the detuning from the exact resonant value of the kicking period \cite{WGF03,PhysRevLett.89.084101,wimberger:084102}. For the quantum resonances, the underlying pseudo-classical model is completely integrable and corresponds in good approximation to the dynamics of a classical pendulum \cite{WGF03,wimberger:084102}.

In this paper we apply well-known semiclassical methods to describe the behavior of fidelity close to the lowest-order quantum resonances of the QKR. We extend previous analytical results at exact resonance \cite{WB2006} to a broader parameter regime, recently measured in experiments performed by Wu and co-workers \cite{fid-har}. The behavior of classical \cite{PhysRevE.68.036212} and quantum fidelity \cite{PhysRevE.68.036216,PhysRevA.71.043803}, in the case when classical motion is integrable, has mainly been addressed numerically so far, while our approach is both numerical and analytical. Also, the recurrences of fidelity found in \cite{PhysRevE.68.036216} for the near-integrable regime of the kicked rotor are just predicted for perturbative variations around small kicking strengths. Our results are more general, allowing, e.g., for strong changes of the fidelity parameter as long as the motion remains nearly resonant. As expected, in the nearly-resonant regime, the temporal behavior of fidelity follows the behavior at exact resonance  the longer, the smaller the detuning from resonance. Indeed,  we show that the exactly resonant result, predicted in \cite{WB2006} by quantum calculations, is retrieved by  pseudo-classical analysis. At large times, however, the exactly resonant fidelity and the nearly resonant one differ, as the latter displays recurrent revivals, while the former steadily decays. Such revivals are approximately periodic. Their period depends on the detuning from resonance and diverges as exact resonance is approached, so this noteworthy phenomenon is unrelated to quantum resonant dynamics. On the other hand, it is quite  unexpected on classical grounds because the system is  chaotic in the proper classical limit. Revivals of fidelity are thus a quantum effect and yet are explained by a (pseudo-)\-classi\-cal analysis that relates them to periodic motion inside pseudo-classical resonant islands. Experimental possibilities to verify our predictions are discussed at the end of the paper.


The dynamics of kicked atoms moving along a line in position space is described, in dimensionless units, by the Hamiltonian \cite{PhysRevA.45,WGF03}:
\begin{equation}
{\cal H}(t)\;=\;\frac{\tau}2p^2\;+\;k\cos(x)\sum\limits_{t'=-\infty}^{+\infty}\delta(t-t')\,,
\end{equation}
where $x$ is the position coordinate and $p$ its conjugate momentum. We use units in which $\hbar=1$ so the parameter $\tau$ plays the role of an effective Planck's constant; $t$ is a continuous time variable, and
$t'$ is an integer which counts the number of kicks. The evolution of the atomic wave function $\psi(x)$ from immediately after one kick to immediately after the next is ruled by the one-period Floquet operator ${\hat U}_{k}\;=\;\exp(-ik\cos({\hat x}))\exp(-i\tau \hat{p}^2/2)$.
Fidelity of the quantum evolution of a state $\psi$ with respect to a change of the parameter $k$ from a value $k_1$ to a value $k_2$ is the function of time $t$ which for all integer $t$ is defined by:
\begin{equation}
\label{fide}
F(k_1,k_2,t)\;=\;\bigl\vert\langle {\hat U}_{k_1}^t\psi |{\hat U}_{k_2}^t\psi\rangle\bigr\vert^2\;.
\end{equation}
Periodicity in space of the kicking potential enforces conservation of
quasi-momentum $\beta$, which is just the fractional part of $p$ thanks to $\hbar=1$. The atomic wave function
decomposes into Bloch waves \cite{PhysRevLett.89.084101,WGF03}, which are eigenfunctions of quasi-momentum,
$\psi(x)=\int_0^1d\beta\;e^{i\beta x}\sqrt{\rho(\beta)}\;\Psi_{\beta}(\theta)$, where $\theta=x\mod(2\pi)$ and
the factor $\rho(\beta)$ is introduced in order to normalize $\Psi_{\beta}$
(it weights the initial population in the Brillouin zone of width one in our units). The dynamics at any fixed value of $\beta$ is formally that of a rotor on a circle, parameterized by the angle coordinate $\theta$ and described by the  wave function $\Psi_{\beta}$.  The Floquet propagator for the rotor is given by
$\hat{{\cal U}}_{\beta,k}\;=\;\exp(-ik\cos(\hat{\theta}))\exp(-i\tau(\hat{{\cal N}}+\beta)^2/2)$, where ${\cal N}=-i\frac d{d\theta}$.
Fidelity (\ref{fide}) may then be written
\begin{equation}
\label{fide1}
F(k_1,k_2,t)\;=\;\biggl\vert\int_0^1d\beta\;\rho(\beta)\langle{\hat{\cal U}}_{\beta,k_1}^{\;t}\Psi_{\beta}
\vert{\hat{\cal U}}_{\beta,k_2}^{\;t}\Psi_{\beta}\rangle\biggr\vert^2,
\end{equation}
so it results from averaging the scalar product under the integral sign over $\beta$ with the weight $\rho(\beta)$.  Note that the rotor's fidelity is the squared modulus of this quantity, so the fidelity (\ref{fide}) of atomic evolution does not coincide with the $\beta$-average of the rotors' fidelities, c.f.\ \cite{WB2006}.
Whenever $\tau=2\pi\ell$ ($\ell$ integer), the evolution is explicitly
solvable \cite{WGF03} and in particular the rotor's fidelity is determined by
\cite{WB2006}
\begin{equation}
\label{fidrot}
\biggl\vert\langle{\hat{\cal U}}_{\beta,k_1}^{\;t}\Psi_{\beta}
\vert{\hat{\cal U}}_{\beta,k_2}^{\;t}\Psi_{\beta}\rangle\biggr\vert^2\;=\;J_0^2(|W_t|\delta k)\;,
\end{equation}
where $J_0$ is the Bessel function of 1st kind and order $0$, $\delta k=k_2-k_1$ and
$|W_t|=|\sin(\pi t\ell(\beta-\tfrac12)\csc(\pi\ell(\beta-\tfrac12)|$. If $2\beta -1$ is an integer
then a so-called QKR resonance occurs and eq.~(\ref{fidrot}) decays in  time proportional to $t^{-1}$. When $\tau$ is close to a resonant value: $\tau=2\pi\ell+\epsilon$, the quantum rotor dynamics may be viewed as the formal quantization of the
pseudo-classical dynamics, defined by the map \cite{WGF03,PhysRevLett.89.084101,wimberger:084102}:
\begin{eqnarray}
\label{epsclass}
I_{t+1}\;&=&\;I_t\;+\;{\tilde k}\sin(\theta_{t+1})\;,\nonumber\\
\theta_{t+1}\;&=&\;\theta_t\;+\;I_t\;+\;\pi\ell\;+\;\tau\beta\mod(2\pi)\;,
\label{map}
\end{eqnarray}
using $\epsilon$ as the Planck's constant, $I=\epsilon{\cal N}$, and ${\tilde k}=\epsilon k$. It is thus possible to investigate the quantum fidelity in the limit of small $\epsilon$ by means of standard methods of semiclassical approximation. In the limit $\epsilon\to 0$, the physical parameter $k$ is fixed, so the pseudo-classical parameter ${\tilde k}\to 0$. As a consequence, for sufficiently small $\epsilon$, the pseudo-classical dynamics (\ref{epsclass}) is in the quasi-integrable regime, even in cases when  the classical kicked rotor dynamics is fully chaotic. It is dominated by the resonant islands at $I_{res}=(2m+\ell)\pi-\tau\beta$, with $m$ integer. As we consider initial atomic states with a narrow distribution of momenta near $p=0$, we may
restrict ourselves to a portion of the pseudo-classical phase space that includes the one island which is located astride $I=0$. We assume $\ell=1$ for simplicity. The pseudo-classical dynamics inside the resonant island is ruled, in continuous time, by the pendulum Hamiltonian \cite{LL92}
$H(\theta, I, {\tilde k})\;=\;\frac12 (I+\bar{\beta})^2\;+\;{\tilde k}\cos(\theta)$,
where $\bar{\beta} \equiv \tau(\beta-\tfrac12)$. We choose $\Psi_{\beta}(\theta)=(2\pi)^{-1/2}$ , so
\begin{equation}
\label{scl}
{\hat{\cal U}}^{\;t}_{\beta,k}\Psi_{\beta}(\theta)\;\sim\;
\frac{1}{\sqrt{2\pi}}\sum\limits_s\left\vert
\frac{
\partial \theta}{\partial\theta'}\right\vert^{-1/2}_{\theta'=\theta'_s}
\;e^{\frac i{\epsilon}\Phi_s(\theta,t)-i\frac {\pi}{2}\nu_s},
\end{equation}
where  $\epsilon>0$ is assumed with no limitation of generality.
The sum is over all trajectories (labeled by the index $s$) which  start with $I=0$ at time $t=0$ and reach position $\theta$ at time $t$. $\theta'=\theta'_s$ are their initial positions, and the function whose derivative is taken in the pre-factor yields $\theta$ at time $t$ as a function of position $\theta'$ at time $0$, given that the initial momentum $I'=0$. Finally, the function $\Phi_s(\theta,t)\;=\;S(\theta,\theta'_s,t)$
is the action of the $s$-th trajectory and $\nu_s$ is the Morse-Maslov index~\cite{HaakeSchulman}. We restrict ourselves to librational motion inside the stable island. The frequency of this motion decreases from $\omega=\sqrt{\tilde k}$ at the island center to $\omega=0$ at the separatrix. For times less than the minimal half-period $\pi/\sqrt{\tilde k}$, there is a single trajectory in eq.~(\ref{scl}). Furthermore,
\begin{eqnarray}
&\frac{\partial\Phi_s(\theta,t)}{\partial t}=\left(\frac{\partial S(\theta,\theta',t)}{\partial
\theta'}\frac{\partial\theta'(\theta,t)}{\partial t}\;+\;\frac{\partial S(\theta,\theta',t)}{\partial t}\right)_{\theta'=\theta'_s}\:=\nonumber\\
&\left(\frac{\partial S(\theta,\theta',t)}{\partial t}\right)_{\theta'=\theta'_s}=-H(\theta_s',0)\;=-\frac{\bar\beta^2}{2}+\;{\tilde k}\cos(\theta_s')\,.
\end{eqnarray}
For $t$ fixed and $\epsilon \to 0$ we use $\theta'(\theta,t)\sim\theta-\bar{\beta }t$ in this equation, so
$\Phi(\theta,t)\sim -\tfrac 12\bar{\beta}^2t\;+\;\tilde{k}\int_0^tdt'\;\cos
(\theta-\bar{\beta} t)= -\tfrac 12\bar{\beta}^2t+2\frac{\tilde k}{\bar\beta}\sin(\tfrac{\bar\beta}2t)\cos(\theta-\tfrac{{\bar\beta}t}2)$.
Replacing all this in eq.~(\ref{fide1}), we find for the rotor's fidelity in the limit when $\epsilon\to 0$ at constant $t$:
\begin{gather}
\label{fidescl}
\biggl\vert\langle{\hat{\cal U}}_{\beta,k_1}^{\;t}\Psi_{\beta}
\vert{\hat{\cal U}}_{\beta,k_2}^{\;t}\Psi_{\beta}\rangle\biggr\vert^2\;\sim\;
\biggl\vert\frac1{2\pi}\int_0^{2\pi}d\theta\;e^{iB(\beta,t)\cos(\theta-\tfrac{{\bar \beta}t}2)}\biggr\vert^2 \nonumber\\ \;=\;J_0^2(B(\beta,t))\;,
\end{gather}
where $B(\beta,t)=2\tfrac{\delta k}{\bar\beta}\sin(\tfrac{{\bar \beta} t}2)$.
Since $B(\beta,t)\approx|W_t|$ in eq.~(\ref{fidrot}) for $\tau=2\pi\ell$ and $\beta\approx\tfrac12$, we see that the pseudo-classical approximation along with the pendulum approximation well reproduce the exact quantum calculation (\ref{fidrot}) when $\epsilon\to 0$ at fixed $t$. 
In the final step of integrating over quasi-momenta to find the fidelity for atoms (as distinct from fidelity for rotors), the pseudo-classical approximation plays no role since the particle's dynamics, unlike the rotor's, does not turn pseudo-classical in the limit $\epsilon\to 0$ \cite{PhysRevLett.89.084101}. Replacing (\ref{fidrot}) in eq. (\ref{fide1}) and computing the integral with a uniform distribution of $\beta$ in $[0,1)$ shows that the complete fidelity (\ref{fide1}) saturates to a non-zero value in the course of time \cite{WB2006}.

Next we address the asymptotic regime where $\epsilon\to 0$ and $t\epsilon^{1/2}\sim$ const.
To this end, the exact solution of the pendulum dynamics is needed in order to compute actions; however, some major features of fidelity are accessible by exploiting the harmonic approximation of the pendulum Hamiltonian. We replace the pendulum by the quadratic Hamiltonian $H(I,\theta)=\tfrac12(I+\bar{\beta})^2+\tfrac{\omega^2}2\theta^2$, where $\omega=\sqrt{\tilde k}$ and a shift of $\theta$ by $\pi$ is understood.   Except at exact multiples of the period, there is one harmonic oscillator trajectory in the sum in (\ref{scl}); moreover, Maslov indices do not depend on the trajectory. Straightforward calculations yield $\theta'(\theta,t)\;=\;\sec(\omega t)\bigl(\theta-{\bar{\beta}}\omega^{-1}\sin(\omega t))$ and
$\Phi(\theta,t)\;=\;\bar{\beta}\theta\;\bigl(\sec(\omega t)-1\bigr) - (\omega^{-1}\bar{\beta}^2+\omega\theta^2)\tan(\omega t)/2$, and so:
\begin{gather}
\langle{\hat{\cal U}}_{\beta,k_1}^{\;t}\Psi_{\beta}
\vert{\hat{\cal U}}_{\beta,k_2}^{\;t}\Psi_{\beta}\rangle\;\sim\;\nonumber\\
\frac{e^{i\lambda(t)}}{2\pi\sqrt{|\cos(\omega_1t)\cos(\omega_2t)|}}\;
\int_{-\pi}^{\pi}d\theta\;e^{\;
\tfrac{i}{2\epsilon}\{A(t)\theta^2+C(t)\bar\beta^2-2\bar\beta\theta B(t)\}
} \;,
\label{step1}
\end{gather}
where
$A(t)=\omega_2\tan(\omega_2t)-\omega_1\tan(\omega_1t), B(t)=\sec(\omega_2t)-\sec(\omega_1t)$ and
$C(t)=\omega_2^{-1}\tan(\omega_2t)-\omega_1^{-1}\tan(\omega_1t)$.
$\lambda(t)$ is a phase factor accumulated by the Maslov indices and it just depends on time, rendering it irrelevant for our present purposes. We next insert eq.~(\ref{step1}) in eq.~(\ref{fide1}) and choose for $\rho(\beta)$ a uniform distribution in some interval $\bigl[\tfrac12-b,\tfrac12+b\bigr)$, with $0 \leq b \leq 1/2$. It is necessary to assume that $b$ is smaller than the halfwidth of the pseudo-classical resonant island, because the harmonic approximation we have used is valid only inside that island. Then
\begin{gather}
F(k_1,k_2,t)\;\sim\;
\frac{1}{16\pi^2
b^2\tau^2\vert\cos{(\omega_1t)}\cos{(\omega_2t)}\vert}\nonumber\\ \times \ \biggl\vert\int_{-\pi}^{\pi}d\theta\;e^{-\tfrac{i}{2\epsilon}\Lambda_1(\theta,\epsilon,t)}
\int_{-\tau b}^{\tau b}d\bar\beta\;e^{-\tfrac{i}{2\epsilon}\Lambda_2(\bar\beta,\theta,\epsilon,t)}\biggr\vert^2 \;,
\label{dide3}
\end{gather}
where $\Lambda_1(\theta,\epsilon,t)=(A(t)-B^2(t)C(t)^{-1})\theta^2$ and 
$\Lambda_2(\bar\beta,\theta,\epsilon,t)=\bigl(\bar\beta\sqrt{C(t)}-B(t)C(t)^{-1/2}\theta\bigr)^2$.
As $\Lambda_2\sim\epsilon^{-1/2}$ in the limit when $\epsilon\to 0$ and $t\sqrt{\epsilon}\sim$const., the limits in the
$\bar\beta$-integral in (\ref{dide3}) may be taken to $\pm\infty$:
$$
\int_{-\tau b}^{\tau b}d\bar\beta\;e^{-\tfrac{i}{2\epsilon}\Lambda_2(\bar\beta,\theta,\epsilon,t)}\;\sim\;
(2\pi)^{1/2}\epsilon^{1/2}C(t)^{-1/2}e^{-i\pi/4}\;.
$$
Due to this approximation, (\ref{fidapproxinf}) below is valid in the asymptotic regime where
$\epsilon$ is small compared to $b^2$.
The remaining $\theta$-integral is dealt with similarly, because the pre-factor of $\theta^2$ in $\Lambda_1$ is $\sim\epsilon^{-1/2}$. Thus finally
\begin{gather}
F(k_1,k_2,t)\;\sim\;\frac{\epsilon^2}
{16\pi^2b^2|C(t)A(t)-B(t)^2||\cos(\omega_1t)\cos(\omega_2t)|} \nonumber\\
\;=\;\frac{\epsilon^2\omega_1\omega_2}{8\pi^2b^2|4\omega_1\omega_2-\omega_+^2\cos(\omega_-t)
-\omega_-^2\cos(\omega_+t)|}\;,
\label{fidapproxinf}
\end{gather}
where $\omega_{\pm}=\omega_1\pm\omega_2$.
Singularities of this expression are artifacts of the approximations used in evaluating the integrals in
\eqref{dide3}, which indeed break down when the divisor in (\ref{fidapproxinf}) is small compared to  $\epsilon$.
However, they account for the periodic ``revivals" that are observed in the fidelity at large times, with the beating period $T_{12}=2\pi/|\omega_-|$ (Fig.~2 (a)). With a quite narrow distribution  of $\beta$, however, fidelity is at long times dominated by the ``resonant" rotors ($\beta=0$ or $\beta=1/2$ respectively), and then revivals occur with the period $T_{12}/2$ (Fig.~1).
Indeed, with the purely resonant $\beta$, eq.~(\ref{dide3}) yields:
\begin{gather}
F(k_1,k_2,t)\;\equiv\;F_{res}(k_1,k_2,t)
\sim  \nonumber\\
\frac{{\epsilon}}{2\pi}\frac{1} {|\omega_2\cos(\omega_1t)\sin(\omega_2t)-\omega_1\cos(\omega_2 t)\sin(\omega_1t)|}\;,
\label{fidapprox}
\end{gather}
which has singularities in time with the mentioned periodicity of $T_{12}/2$.
This behavior of resonant rotors has a simple qualitative explanation. As the initial state of the rotor corresponds to momentum $I=0$, at that value of quasi-momentum ($\beta=\tfrac12$) the stationary-phase
trajectories of the two harmonic oscillators, which were started at $I=0$, exactly return to $I=0$ whenever time is a multiple of the half-period $T_{12}/2$, and so fully contribute to fidelity, in spite of their angles being different by $\pi$ in the case of odd multiples. At $\beta\neq 0$ this symmetry is lost.
\begin{figure}[t]
\centerline{\epsfig{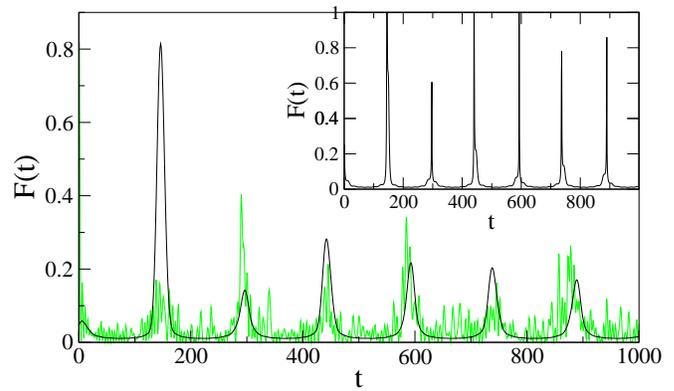}}
\caption{(color online)
Fidelity as predicted by eq.~\eqref{fidapprox} -- because of the singularities of the analytical formula the curve is folded with normalized Gaussians with a standard deviation of $t \approx 6$ kicks ({\it solid black line}) -- and numerical data ({\it grey/green curve}), for $k_1 = 0.8\pi$, $k_2 = 0.6\pi$ and detuning $\epsilon = 0.01$ from $\tau-\epsilon=2\pi$. In the inset, the non-smoothed result \eqref{fidapprox} is shown. 
}
\label{fig:1}
\end{figure}
\begin{figure}[t]
\centerline{\epsfig{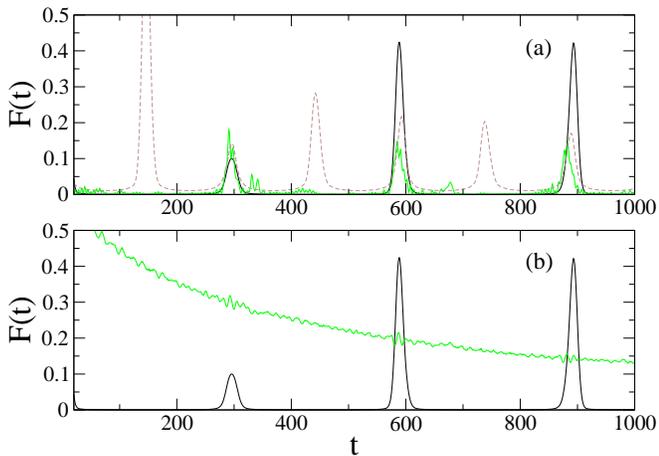}}
\caption{(color online)
Same as in Fig.~\ref{fig:1} for an ensemble of 5000 equidistantly chosen rotors ({\it solid grey/green lines}) with a width of (a) $\Delta\beta = 0.05$ (or $\Delta\bar\beta \approx 0.31$) around the resonant value, covering half the width of the resonance island in the phase space induced by \eqref{map}, and (b) $\Delta\beta = 1$, covering the full phase space, compared with the smoothed (see caption of Fig.~\ref{fig:1}) version of eq.~\eqref{fidapproxinf} ({\it solid black lines}). In (a) the intermediate revival peaks observed in Fig.~\ref{fig:1} disappear as predicted by \eqref{fidapproxinf}. The dashed line in (a) reproduces the smoothed analytic formula from Fig.~\ref{fig:1}. For $\beta$ distributed over the full Brillouin zone in (b), the revivals are barely visible since the average includes many nonresonant rotors performing rotational motion in phase space, which is not decribed by our theory valid just for the librational island motion.
}
\label{fig:2}
\end{figure}
Comparing numerical data (obtained by repeated application of the Floquet operator to the initial wavefunction) with the analytical predictions we find excellent agreement. We observe the expected peak structure of the revivals in Fig.~\ref{fig:1} and the loss of intermediate revival peaks at $T_{12}/2$ in Fig.~\ref{fig:2}(a). The time scale on which the revivals occur is proportional to $\epsilon^{-1/2}$ and of crucial impact to experimental measurements: conservation of coherence has been shown for up to 150 kicks (see \cite{PhysRevLett.90.124102}) with cold atoms, making an observation of the revivals for reasonable $\epsilon \lesssim 0.01$ possible. Earlier realizations of the QKR were implemented using cold atoms \cite{fid-ox,OC.179.137,PhysRevLett.94.174103} with broad distributions in quasi-momentum. Nowadays, much better control of quasi-momentum is provided by using Bose-Einstein condensates (see \cite{PhysRevLett.96.160403,PhysRevE.70.056206}), which allows for a restriction in $\beta$ up to 0.2 $\%$ (as achieved in \cite{PhysRevLett.96.160403}) of the Brillouin zone. This would allow to verify our results by conveniently reducing the intervals in quasi-momentum and thus retracing the revivals with period $T_{12}/2$ to the exactly resonant and the revivals with period $T_{12}$ to the near-resonant rotors. There exists an interesting second possibility to measure the transition from eq.~(\ref{fidapprox}) to eq.~(\ref{fidapproxinf}) with just cold atoms, since the $\bar{\beta}$ we use scales with the kicking period, i.e. $\bar{\beta} = \tau (\beta - 1/2)$. Due to this scaling, the limit $\tau \rightarrow 0$ (automatically implying also $\epsilon \to 0$, c.f. \cite{PhysRevLett.94.174103}) permits a measurement of eq.~\eqref{fidapprox}, even with an ensemble of cold atoms whose quasi-momenta occupy the full Brillouin zone. Also the momentum selective interferometric measurements of fidelity \cite{fid-har} allow to select narrow intervals of quasi-momenta, and hence would permit to check our predictions experimentally.

To summarize, we predict fidelity revivals in the QKR close to quantum resonance using a semiclassical ansatz. Our results are supported by numerical data showing the same characteristic revival peaks. Every second peak vanishes once the symmetry of the initial quasi-momentum distribution on the resonance island is broken. This makes for a surprising transition that could be measured with both cold and ultracold atoms owing to the scaling of $\bar{\beta}$ or the use of momentum selective methods, as described in the previous paragraph. 

Support by the Excellence Initiative through the Global Networks Mobility Measures and the Heidelberg Graduate School of Fundamental Physics (DFG grant GSC 129/1), and by a Short Visit Grant (DAAD) is acknowledged.

\end{document}